\documentclass[aps,prl,twocolumn,showpacs,nobibnotes]{revtex4-1}

\usepackage{graphicx}
\usepackage{subfigure}

\usepackage{amsmath}
\usepackage{amssymb}
\usepackage{amsthm}

\usepackage{color}

\begin{document}

\title{Applying the coupled cluster ansatz to solids and surfaces
in the thermodynamic limit} 

\author{Thomas~Gruber}
\affiliation{%
Max Planck Institute for Solid State Research, Heisenbergstrasse 1, 70569 Stuttgart, Germany}
\author{Ke~Liao}
\affiliation{%
Max Planck Institute for Solid State Research, Heisenbergstrasse 1, 70569 Stuttgart, Germany}
\author{Felix~Hummel}
\affiliation{Institute for Theoretical Physics, Vienna University of Technology, Wiedner Hauptstrasse 8-10, 1040 Vienna, Austria}
\author{Theodoros~Tsatsoulis}
\affiliation{Max Planck Institute for Solid State Research, Heisenbergstrasse 1, 70569 Stuttgart, Germany}
\author{Andreas~Gr\"uneis}
\email{andreas.grueneis@tuwien.ac.at}
\affiliation{Institute for Theoretical Physics, Vienna University of Technology, Wiedner Hauptstrasse 8-10, 1040 Vienna, Austria}
\affiliation{Max Planck Institute for Solid State Research, Heisenbergstrasse 1, 70569 Stuttgart, Germany}

\begin{abstract} 
Modern electronic structure theories can predict and simulate a wealth of phenomena
in surface science and solid state physics.
In order to allow for a direct comparison with experiment such $ab-initio$ predictions
have to be made in the thermodynamic limit, increasing the computational cost of many-electron wave function theories substantially.
Here, we present a method that achieves thermodynamic limit results for solids and surfaces using the `gold standard' coupled-cluster
ansatz of quantum chemistry with unprecedented efficiency. We study the energy difference between carbon diamond
and graphite crystals, adsorption energies of water on $h$-BN as well as the cohesive energy of the Ne solid, demonstrating the
increased efficiency and accuracy of coupled cluster theory for solids and surfaces.

\end{abstract}

\maketitle

\section{Introduction}

Modern {\it ab-initio} methods to solve the electronic Schr\"odinger equation for real solids
and molecules such as density functional theory or wave function based methods are becoming increasingly
accurate and efficient~\cite{Sun2016,Foulkes2001,Yang640,Booth2013}.
However, in contrast to molecular systems, properties of solids and surfaces need to
be calculated in the thermodynamic limit. The convergence
towards the thermodynamic limit with respect to the number of particles is very slow,
often exceeding the computational resources of even modern super computers.
This is particularly the case for many-electron wave function based theories that
allow for a systematic improvability upon the description of the electronic correlation
energy.
Nonetheless these methods are becoming increasingly popular in
theoretical physics as well as chemistry to treat electronic correlation
in periodic condensed matter systems with high accuracy
\cite{Yang640,Booth2013,Nolan2009,Paulus1999,Shepherd2014,PhysRevB.89.205138,
Hermann2008,Schwerdtfeger2010-ir,McClain2017,PhysRevX.7.031059,PhysRevX.5.041041}.

Electronic correlation is for the most part a short-ranged phenomenon.
The proper description of the wave function shape at short
interelectronic distances allows for capturing the largest fraction of the
correlation energy in solids~\cite{Foulkes2001,Gruneis2013}.
Significant progress has been achieved for many-electron wave function based theories by
exploiting the locality of electronic correlation in large molecules and solids. 
The development of so-called local correlation methods and embedding theories
has improved their computational efficiency
considerably~\cite{Haettig2012a,doi:10.1063/1.4773581,doi:10.1063/1.3641642,Usvyat2015,PhysRevLett.109.186404,PhysRevB.89.035140,doi:10.1021/ct300544e,Zgid2014,Carter1999,Manby2012}.
However, theories that approximate long-range correlation effects such as van der Waals interactions
must carefully be checked for convergence with respect to the employed cutoff parameters to allow for
accurate and predictive {\it ab-initio} studies of real materials.
This is of particular importance in condensed matter systems where the accumulation
of weak van der Waals interactions can become a non-negligible contribution to the property of interest
as for example in the case of the energy difference between carbon diamond and graphite or
the adsorption of a water molecule on an $h$-BN sheet.
Pairwise additive interatomic van der Waals interactions cause an $1/N$ convergence of
the electronic correlation energy per unit cell in insulating three dimensional systems, where $N$ is the number of
explicitly correlated atoms.
However, in general the exact form of these scaling laws depends on the dimensionality and electronic response properties
of the system, e.g. the adsorption energy of molecules on two dimensional insulating surfaces exhibits an $1/N^2$ convergence.
Moreover we note that collective phenomena such as plasmons in metallic systems can also modify
the observed scaling laws~\cite{PhysRevLett.96.073201}.
For these reasons robust and reliable approximations to long-range correlation effects are non-trivial.

Many-body methods such as coupled cluster or configuration
interaction theory can describe both long- and short-ranged electronic
correlation effects with high accuracy.
However, the scaling of the computational complexity of these theories with respect
to system size is either of a high-order polynomial or even exponential form.
Therefore it is difficult to treat long-range correlation effects
in a computationally efficient manner using these theories.
This has lead to the development of various techniques that partition the correlation problem
according to a predefined criterion such as the distance between electron pairs or fragment size.
Local theories employ correlation energy expressions that depend on localized electron pairs,
making it possible to treat long-distance pairs using computationally more efficient yet less accurate theories.
Embedding theories typically aim at combining the computational efficiency of mean field theories for
the long-range with the high accuracy of wave function based methods applied to small fragments only.
In this work we introduce an efficient method that seamlessly integrates
long-range correlation effects for solids without any predefined criteria
such as cutoff distance or fragment size.
Our approach is inspired by structure factor interpolation techniques as performed in the field
of Quantum Monte Carlo theory~\cite{PhysRevLett.97.076404}.
However, in coupled cluster theory the structure factor,
being the functional derivative of the total energy with respect to the
Coulomb kernel, is not directly available. Instead we seek to interpolate
the partial functional derivative of the coupled cluster correlation energy
expression with respect to the Coulomb kernel.
The interpolation scheme is chosen such that it is directly transferable to
systems with arbitrary dimensions including solids and surfaces.
Due to the adverse scaling of the computational complexity in coupled cluster theories, the proposed method allows for
reducing the computational cost by several orders of magnitude without
compromising accuracy compared to previous studies~\cite{Booth2013}.

\section{Theory}
The electronic correlation energy can be calculated in a plane wave basis set
using the following expression~\cite{liao2016}
\begin{equation}
E_c=\langle \Psi_0 | H-E_0 | \Psi \rangle = {\sum_{{\bf G}}}' {v}({\bf G}) {S}({\bf G}).
\label{eq:ecorrg}
\end{equation}
In the above equation ${\bf G}$ corresponds to a plane wave vector that is
defined as ${\bf G}={\bf g} + {\Delta \bf k}$, where ${\bf g}$ is a reciprocal lattice vector
and ${\Delta \bf k}$ is the difference between any two Bloch wave vectors that are conventionally
chosen to sample the first Brillouin zone.
${v}({\bf G})$ is the Coulomb kernel
in reciprocal space that diverges at ${\bf G}=0$, making it numerically necessary to disregard this contribution
to the sum as indicated by the apostrophe.
Thus, ${S}({\bf G})$ is the partial functional derivative of the correlation energy with respect to
${v}({\bf G})$ and we will return to its explicit definition later as well as in the supplementary information~\cite{liao2016}.

The thermodynamic limit is approached as $N\rightarrow \infty$, where $N$ is the number of particles in
the simulation cell while the density is kept constant.
Finite size errors are defined as the difference between the thermodynamic limit and the
finite simulation cell results.
For electronic correlation energies obtained using many-electron perturbation theories
these errors typically decay as $1/N$ as a consequence of long-range interatomic van der Waals forces.
In the thermodynamic limit $\sum_{{\bf G}}$ of Eq.~(\ref{eq:ecorrg}) is replaced by $\int_{{\bf G}}$.
Therefore finite size errors in the correlation energy of periodic systems
originate from two sources~\cite{Holzmann2016}:
(i) quadrature errors in the summation over ${\bf G}$, and
(ii) the slow convergence of ${S}({\bf G})$ with respect to the employed
supercell size or $k$-point mesh.
In the following we will discuss how to reduce both errors substantially.

We first seek to discuss finite size errors originating from the quadrature
in the summation over ${\bf G}$.
These contributions can be partitioned into the ${\bf G}=0$ volume element contribution and the remaining
terms.
We note that as a result of the Coulomb divergence, the integrable contribution of
$S({\bf 0})v({\bf 0})$ to the correlation energy is usually neglected in computer
implementations of Eq.~(\ref{eq:ecorrg})~\cite{liao2016,McClain2017}.
However, this is the dominant contribution to the finite size error of the correlation energy
of insulators. A Taylor expansion of  ${S}({\bf G})$ around ${\bf G}=0$ shows
that ${S}({\bf G})$ exhibits a quadratic behavior close to zero, explaining the $1/N$ decay of the
finite size error for three dimensional insulators~\cite{liao2016}.
An estimate of $S({\bf 0})v({\bf 0})$ can be obtained by spherically
averaging ${S}({\bf G})$ and interpolating around ${\bf G}=0$. Subsequently, the interpolated
function is multiplied with the analytic Coulomb kernel and integrated over
a sphere around ${\bf G}=0$, yielding an estimate of ${S}({\bf 0}){v}({\bf 0})$~\cite{liao2016}.
However, this approach is not well defined for anisotropic systems because
it requires a spherical cutoff parameter.
In this work we propose to interpolate ${S}({\bf G})$ using a tricubic interpolation without spherical averaging.
Once obtained the interpolation of ${S}({\bf G})$ and the analytic expression for the Coulomb kernel
allows for integrating over ${\bf G}$ on a very fine grid, simulating the thermodynamic limit integration.
This approach accounts for the ${S}({\bf 0}){v}({\bf 0})$
contribution to the correlation energy and reduces quadrature errors originating from too coarse a
Brillouin zone sampling. We will refer to coupled cluster correlation energies obtained using
this interpolation strategy as CC-FS.

\begin{figure}
    \begin{center}
        \includegraphics[width=8.0cm]{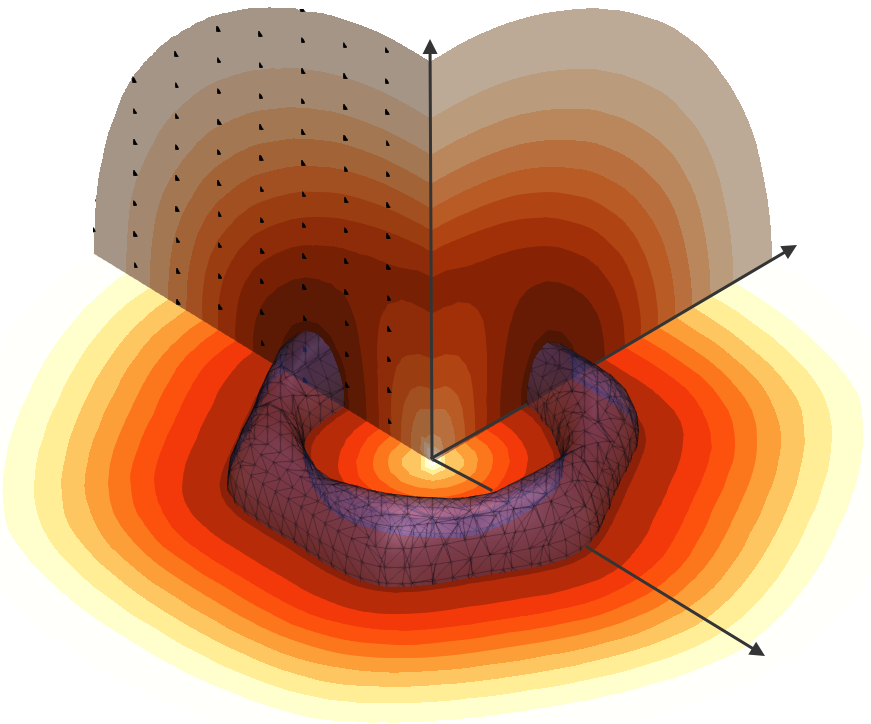}
    \end{center}
    \caption{ 
{Partial derivative of correlation energy with respect to Coulomb kernel for graphite.}
Slices and isosurface of $S({\bf G})$ for carbon graphite in the ABC stacking.
Darker colors indicate more negative values. White corresponds to zero.
The blue isosurface is a hexagonally shaped torus and reflects the anisotropic shape for $S({\bf G})$.
Black dots represent the sampling points using a $4\times4\times4$ $k$-point mesh.
    }
    \label{fig:1}
\end{figure}

To illustrate the importance of the interpolation method we consider the following example.
Figure~\ref{fig:1} shows slices and an isosurface of the interpolated $S({\bf G})$
for carbon graphite in the ABC stacking. Black dots indicate sampling points of  $S({\bf G})$ obtained using
coupled cluster singles and doubles theory and a $4\times4\times4$ $k$-point mesh.
This figure illustrates that $S({\bf G})$ is very anisotropic
around ${\bf G}=0$. Furthermore we show that even a $4\times4\times4$ $k$-point mesh sampling
indicated by the black dots corresponds to a relatively coarse grid, causing non-negligible quadrature errors.
We will return to the discussion of the results for the correlation energy later.

We now turn to the discussion of finite size errors in semiconductors and metals.
We stress that small gap systems suffer from
a relatively slow convergence of  ${S}({\bf G})$ with respect
to the studied system size.
This behavior can be understood by considering the definition of 
${S}({\bf G})$ in second-order M\o ller-Plesset perturbation (MP2) theory
\begin{equation}
{S}({\bf G})=\sum_{{\bf k}_i,{\bf k}_j,\\ {\bf k}_a}\sum_{n_i, n_j,\\ n_a,n_b}
\frac{\Gamma_{ij}^{ab}({\bf G})}{\epsilon_i+\epsilon_j-\epsilon_a-\epsilon_b},
\label{eq:TofG}
\end{equation}
where
$\epsilon_i$ correspond to one-electron energies usually obtained from Hartree--Fock theory.
The indices $i,j$ and $a,b$ label occupied and virtual orbitals respectively and
are understood to be a shorthand for the Bloch wave vector ${\bf k}_i$ and
a band index $n_i$. 
Due to momentum conservation
${\bf k}_b$ can be calculated from the other Bloch wave vectors in the above equation.
$\Gamma_{ij}^{ab}({\bf G})$ is defined in the supplementary information.
The summation over Bloch vectors in Eq.~(\ref{eq:TofG}) introduces quadrature errors that
cause the slow convergence of ${S}({\bf G})$ towards the thermodynamic limit.
In the case semiconductors or metals these errors can become significant
because $1/(\epsilon_i+\epsilon_j-\epsilon_a-\epsilon_b)$ varies strongly depending on
${\bf k}_i,{\bf k}_j,{\bf k}_a$ and ${\bf k}_b$.
In particular materials with a Dirac cone at the Fermi surface such as
graphene exhibit a large variation of the denominator between zero and several eV depending on ${\bf k}$.
As a result Eq.~(\ref{eq:TofG}) needs to be calculated using a finer $k$-point mesh to reduce
quadrature errors.
We will show in this work that the above quadrature errors can be
substantially reduced by calculating and averaging ${S}({\bf G})$ for a set of shifted
$k$-point meshes.
Note that the vectors ${\bf G}$ are not affected by shifting the employed $k$-mesh because ${\bf G}$
depends only on the difference between any two Bloch wave vectors $\Delta {\bf k}$.
We replace ${S}({\bf G})$ in Eq.~(\ref{eq:ecorrg}) with an average obtained
for $N_t$ different $k$-meshes shifted from $\Gamma$ by ${\bf t}_i$ such that
\begin{equation}
\bar{S}({\bf G})=\frac{1}{N_t}\sum_{i=1}^{N_t}{S}_{{\bf t}_i}({\bf G}).
\label{eq:tsavg}
\end{equation}
The shifts ${\bf t}_i$ are chosen such that they sample the first Brillouin zone
uniformly. Coupled cluster theory calculations for different shifts can be performed independently from
each other and the computational complexity scales only linearly with respect to $N_t$.
Coupled cluster theory energies that have been obtained using this twist-averaging
technique will be referred to as CC-TA or CC-TA-FS if the interpolation method has been
employed as well.

We note that quantum Monte Carlo (QMC) methods
employ finite size corrections that share similarities with the methods outlined above~\cite{Lin2001,Filippi1999,Holzmann2016}.
However, QMC methods such as diffusion Monte Carlo are real-space theories that provide estimates of total energies
rather than partitioning the energy into
a Hartree--Fock and an electronic correlation contribution.
An advantage of the partitioning ansatz is that Hartree--Fock energy contributions
can be converged to the thermodynamic limit independently from the correlation energy at little extra computational cost.
Consequently, finite size corrections are only required for the comparatively smaller correlation energy contributions.
In passing we note that auxiliary field quantum Monte Carlo theory employs finite size corrections that
are based on parametrized density functionals obtained from finite uniform electron gas simulation
cells~\cite{PhysRevLett.100.126404}. 
Such corrections work for solids but have not yet been applied to surfaces or molecular crystals where
they are expected to be less accurate.


\section{Results}
We now turn to the discussion of the results obtained using the methods
outlined above.
The present computations were performed using the \texttt{VASP} code~\cite{Kresse1994,Kresse1996-te}
and the projector augmented wave method~\cite{Blochl1994}. The coupled cluster theory
calculations were partly performed using the newly developed \texttt{cc4s}
code~\footnote{The details about the \texttt{cc4s} code will be elaborated in a future paper.}
interfaced with \texttt{VASP} and employing the automated tensor contraction engine \texttt{CTF}~\cite{Solomonik2014}.
More technical details are outlined in the supplementary information.

\begin{figure}[h]
 \includegraphics[width=7.0cm,clip=true]{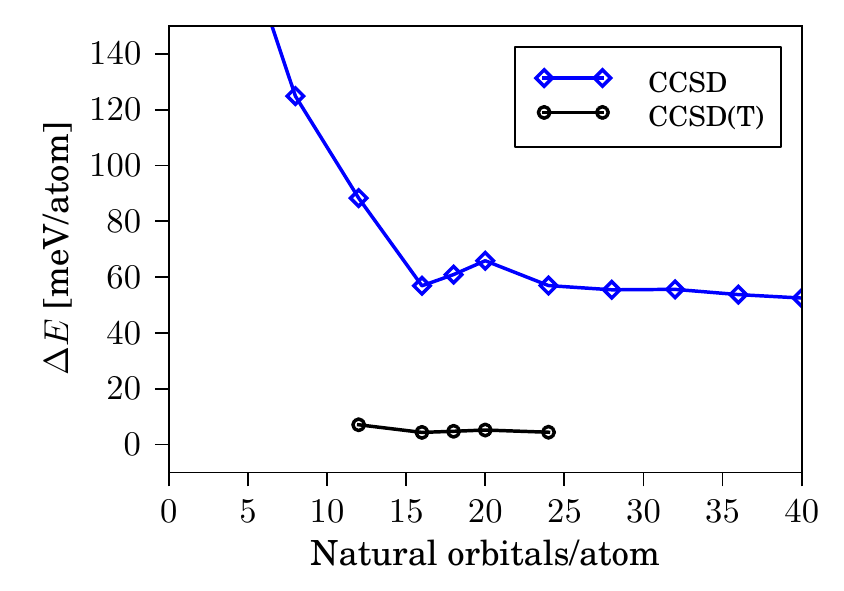}
\caption{Convergence of the energy difference between the carbon diamond and graphite using CCSD
and 2 atomic unit cells with respect to the number of natural orbitals used. The (T) correlation
energy contribution has been added to the CCSD energy using 16 natural orbitals per atom.
\label{fig:S1}}
\end{figure} 
\begin{figure}
    \begin{center}
        \includegraphics[width=7.0cm,clip=true]{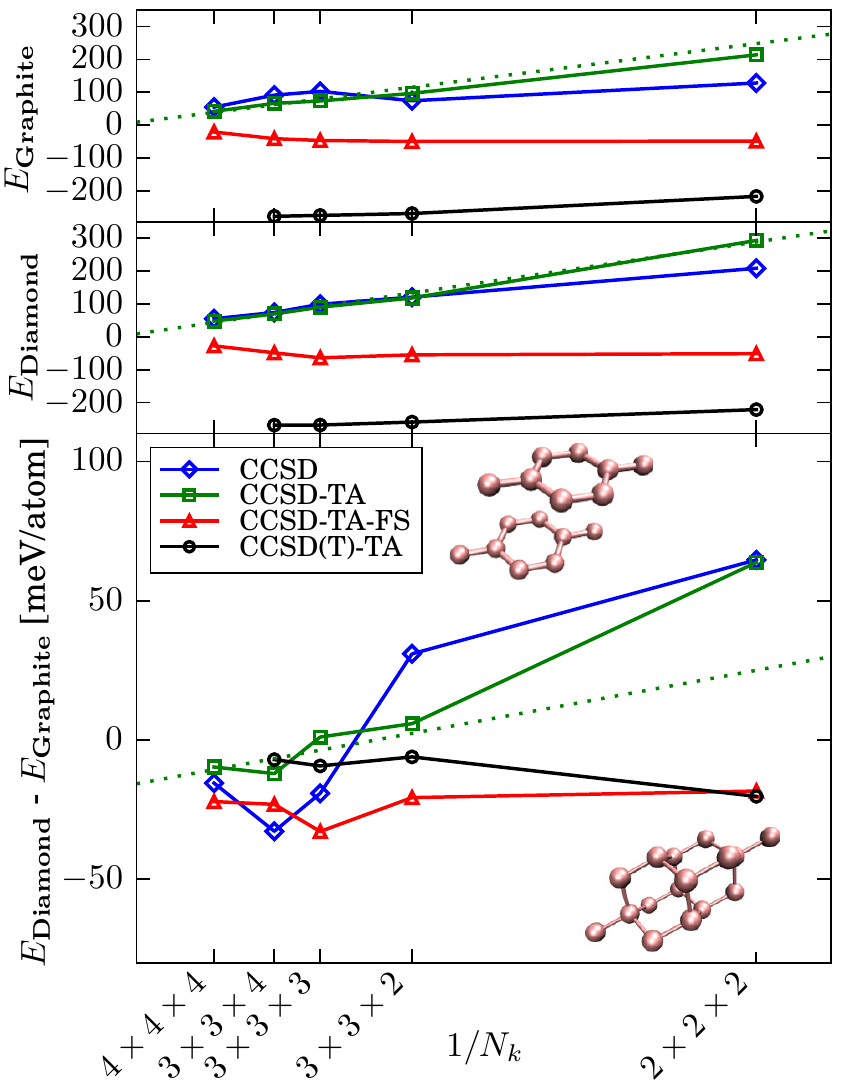}
    \end{center}
    \caption{
{Energy difference between carbon diamond and graphite phases.}
The dotted line represents the linear fit of the uncorrected values with shifts.
(T) corrections (black) are on top of CCSD-TA-FS energy obtained using a $4\times4\times4$ $k$-point mesh (red).
Zero point energies are included and stabilize graphite compared to diamond by 9~meV/atom.
    }
    \label{fig:2}
\end{figure}

As a first application
we investigate  the carbon diamond and graphite crystals.
Before discussing the thermodynamic limit convergence we seek to 
address the convergence of the calculated correlation energy differences
with respect to the employed orbital basis.
We employ MP2 natural orbitals that are obtained using a procedure outlined in Ref.~\cite{Gruneis2011a}.
Figure~\ref{fig:S1} shows the convergence of coupled cluster singles and doubles (CCSD)  and perturbative triples (T) correlation energy
differences with respect to the number of bands using a $2\times2\times2$ $k$-point mesh, respectively.
We find that calculations using 16 orbitals per carbon atom yield an energy difference that
agrees to within 4~meV/atom compared to results obtained using 40 natural orbitals per atom.
The (T) correction to CCSD converges even faster with respect to the number of orbitals
and is fortuitously close to zero in the case of the $2\times2\times2$ $k$-point mesh.
We stress that natural orbitals allow for a significantly more systematic truncatability and improved basis
set incompleteness error cancellation between different systems compared to virtual Hartree--Fock or
density functional theory orbitals.
Achieving the same level of accuracy requires several hundred virtual Hartree--Fock orbitals per atom.
Convergence with respect to other computational parameters has been checked and is
discussed in the supplementary information.\\
We now turn to the discussion of finite size errors in total correlation energies.
The top and middle panel in Figure~\ref{fig:2} show CCSD
correlation energies retrieved as a function of the number of $k$-points of graphite and diamond, respectively.
We note that twist averaging (TA) is necessary for CCSD correlation energies to achieve a smooth $1/N$
convergence to the thermodynamic limit in particular for graphite. Accounting for quadrature errors by
means of the tricubic interpolation method (CCSD-TA-FS) yields
rapidly convergent correlation energies for both carbon diamond and graphite.
CCSD-TA-FS correlation energies obtained using a $2\times2\times2$ $k$-mesh only deviate
from extrapolated CCSD thermodynamic limit energies by approximately 60~meV/atom. We note that correlation energies
obtained using the same $k$-mesh and CCSD-TA theory exhibit finite size errors on the scale of 200--300~meV/atom.
The CCSD(T)  correlation energies are obtained using twist averaging for the (T) contribution and adding the correction
to the CCSD-TA-FS result obtained using a $4\times4\times4$ $k$-point mesh.
This allows for  investigating the finite size errors of the (T) correction independently from finite size errors of CCSD theory.
We find that (T)
converges rapidly with respect to the employed $k$-mesh size, reflecting its short-rangedness.

The bottom panel in Figure~\ref{fig:2} shows the differences of the total energies of both carbon allotropes including
zero point corrections retrieved
as a function of the employed $k$-point mesh.
Results obtained using CCSD theory without finite size corrections
are depicted by the blue line and  oscillate strongly with increasing $k$-point mesh density.
Using too coarse $k$-point meshes predicts graphite to be more stable than diamond,
whereas denser $k$-point meshes predict diamond to be the more stable allotrope.
Employing the averaging over different shifts yields CCSD-TA results
that converge significantly smoother with increasing  $k$-point mesh density
as shown by the green line.
Furthermore performing the newly proposed tricubic interpolation and integration in addition
to the twist averaging referred to as CCSD-TA-FS yields rapidly convergent energy differences
shown by the red line.
We note that CCSD-TA-FS using a $2\times2\times2$ $k$-point mesh is as close
to the thermodynamic limit as CCSD-TA using a  $4\times4\times4$ $k$-point mesh.
Since the computational complexity scales at least as $\mathcal{O}(N_k^4)$ with respect to the number
of $k$-points this corresponds to a reduction in the computational cost by three orders
of magnitude.
From these calculations we conclude that CCSD theory predicts diamond to be more stable than graphite
by 22~meV/atom including zero point energies.
We have also performed perturbative triples calculations
and added the corresponding correlation energy correction to
our CCSD findings.
CCSD(T) theory  predicts graphite to be slightly less stable than diamond by 7~meV/atom including zero point corrections.
We note that our CCSD(T) results agree with experimental findings to within the observed
precision of CCSD(T) for similar applications which is generally better than 1~kcal/mol (43~meV/atom).
The difference in the experimental Gibbs free energy of carbon diamond and graphite 
at room temperature has been reported to be 25~meV/atom~\cite{WagmanJoRotNBoS1945}, predicting graphite to be more stable than diamond.

\begin{table}[t]
  \caption{{Cohesive energies of solid Neon obtained using MP2, CCSD and CCSD(T) theory.}
  The summarized results have been extrapolated to the complete basis set limit using
  pseudized aug-cc-pV(D,T)Z basis sets and  corrected for basis set superposition errors using
  counterpoise corrections. All units in meV/atom. \label{tab:neon}
  }
\begin{ruledtabular}
    \begin{footnotesize}
    \begin{tabular}{lccccc}
        $k$-mesh        &  { MP2 } & { MP2-FS} & { CCSD} & { CCSD-FS} & { CCSD(T)-FS }    \\ \hline
     $2\times2\times2$  &   -5  &   25   &   4  & 36      &   47        \\
     $3\times3\times3$  &    13 &   17   &  16  & 21      &   32        \\
     $4\times4\times4$  &    17 &   17   &  19  & 19      &   30        \\
     Ref.~\cite{Schwerdtfeger2010-ir} &       &    19  &      &  22     &   27        \\
    \end{tabular}
    \end{footnotesize}
\end{ruledtabular}
\end{table}

Having demonstrated the ability of the proposed method to correct for finite size errors on the scale
of about 100~meV/atom, we now seek to study the thermodynamic limit of the cohesive energy of the weakly bound Neon solid.
In this case we need to correct for finite size errors on the scale of a few meV/atom.
The dominant contribution to the attractive long range interatomic interaction of Neon atoms originates from van der Waals forces.
Table~\ref{tab:neon} summarizes MP2, CCSD and CCSD(T) cohesive energies obtained with and without the proposed
finite size correction. The correction yields MP2 and CCSD cohesive energies using $3\times3\times3$ $k$-meshes that deviate from the
thermodynamic limit results by approximately 1-2~meV/atom, whereas the uncorrected estimates deviate by 2--4~meV/atom. Although the finite
size errors are small on an absolute scale, we stress that the corresponding relative finite size errors of the cohesive energy
are non-negligible.
Our best estimates for the MP2, CCSD and CCSD(T) cohesive energies using finite size corrections and a $4\times4\times4$ $k$-mesh
agree with results obtained using the incremental method to within 3~meV/atom~\cite{Schwerdtfeger2010-ir,Paulus1999}.
Furthermore CCSD(T) predicts a cohesive energy of 30~meV/atom which is in good agreement
with experimental estimates of 27~meV/atom corrected for zero point fluctuations~\cite{Paulus1999}.

\begin{figure}
    \begin{center}
        \includegraphics[width=8.0cm,clip=true]{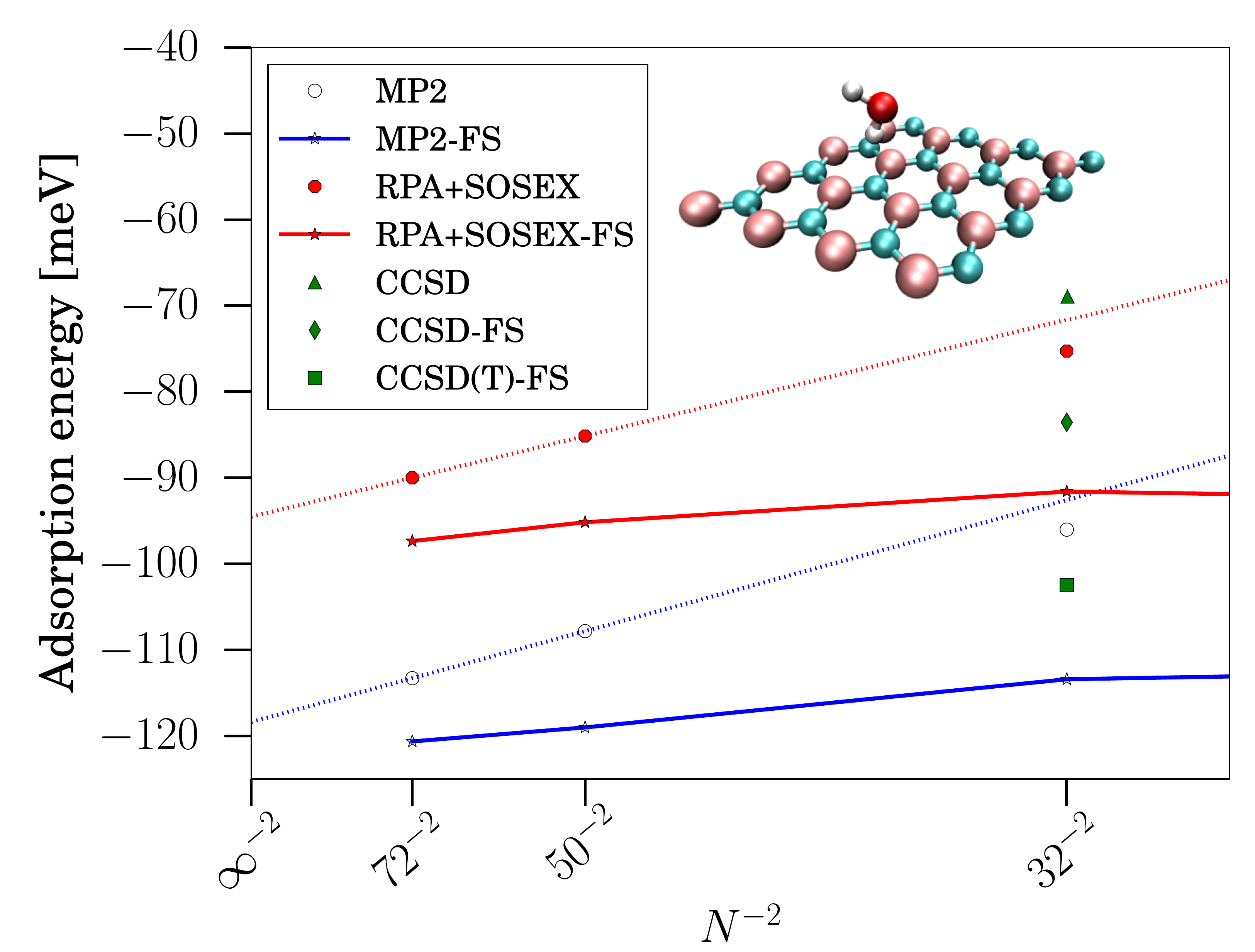}
    \end{center}
    \caption{
{Adsorption energy of a single water molecule on an $h$-BN sheet.}
The energies are
retrieved as a function of atoms in the sheet using MP2, RPA+SOSEX, CCSD and CCSD(T) theory.
FS indicates that finite size corrections are included.
The results have been obtained using
pseudized aug-cc-pVTZ basis sets and  corrected for basis set superposition errors using
counterpoise corrections.
    }
    \label{fig:4}
\end{figure}

As a final demonstration of the applicability of the proposed
method to reach the thermodynamic limit
we study the adsorption energy of a single
water molecule on an $h$-BN sheet. The same system has recently been studied using
diffusion Monte Carlo (DMC), the random-phase approximation (RPA) and dispersion
functionals~\cite{yasmine2017,wu2016}, as well as
molecular MP2~\cite{wu2015} and periodic coupled-cluster theory~\cite{hummel2017},
demonstrating the need for reliable methods that can account
for long-range van der Waals interactions, also to provide benchmark
data.
Furthermore, the recent work of Al-Hamdani
\textit{et al.}~\cite{yasmine2017} illustrates the importance of
long-range correlation effects that account for approximately 25\% of
the reference adsorption energy computed in a $(4\times4)$ unit cell of $h$-BN.
Figure~\ref{fig:4} shows calculated adsorption energies at the level of RPA plus second-order
screened exchange, MP2, CCSD and
CCSD(T) theories retrieved as a function of the number of atoms in the $h$-BN sheet.
Convergence with respect to other computational parameters has been checked and is discussed in
the supplementary information.
Using MP2 theory it is possible to
study very large systems~\cite{yasmine2017} and we find that the MP2
adsorption energy converges slowly to a thermodynamic limit value of
119~meV.
We note that finite size errors for adsorption energies on two
dimensional insulators are expected to decay as $1/N^2$, which is the
predicted scaling from pairwise additive van der Waals interactions~\cite{yasmine2017}.
Applying the proposed finite size correction to MP2 theory for the
$(4\times4)$ unit cell $h$-BN
sheet with 32 atoms
yields an adsorption energy of 113~meV in close agreement with the thermodynamic
limit result.
We observe a similar speed-up in convergence using RPA+SOSEX-FS theory, illustrating the transferability
of the proposed method.
The water adsorption on the 32-atom $h$-BN sheet can also be studied using
the more sophisticated CCSD theory~\cite{hummel2017}.
CCSD with and without finite size corrections yields an adsorption energy of 83~meV and 68~meV,
respectively. The finite size correction of MP2 and CCSD theory agree to within a few meV.
However, we note that CCSD theory underbinds the water molecule.
We estimate the (T) contribution using the 18-atom cell only and find that
CCSD(T) theory yields an adsorption energy of 102~meV and 87~meV with and without finite size correction,
respectively.
The DMC adsorption energy was reported to be 84~meV without finite size corrections and agrees well with
our CCSD(T) results using the same 32-atom $h$-BN sheet disregarding finite size corrections.

\section{Conclusion and Outlook} In conclusion, we have introduced an 
efficient and accurate thermodynamic limit correction for wave function based
theory calculations of solids and surfaces that is free of adjustable parameters
and easy to implement.
We have demonstrated that this correction allows for reducing the computational cost
by several orders of magnitude without compromising accuracy.
We have studied ground state problems where the convergence to the thermodynamic limit is crucial
and finite size errors span a range of 1--100~meV/atom.
Despite the local character of electronic correlation we stress that a proper treatment of long-range
correlation effects is of paramount importance for reliable and highly accurate many-electron theories
in condensed matter systems.
We have applied the proposed finite size correction in combination with the gold standard of quantum
chemistry CCSD(T) theory to calculate the cohesive energy of the Ne solid,
the energy difference between carbon diamond and graphite crystals as well as the adsorption
of a water molecule on an $h$-BN sheet.
In general our CCSD(T) results are in good agreement with experimental findings and DMC results.
This paves the way for a routine use of highly accurate coupled cluster theories
in the field of surface science and solid state physics.
We believe that the ability to predict accurate benchmark results will help
the entire electronic structure theory community to improve further upon computationally
efficient $ab-initio$ theories and to help interpret experimental findings more reliably.
To expand the scope of the proposed techniques even further we will aim at combining them with
explicit correlation and low rank factorization methods~\cite{hummel2017,PhysRevLett.115.066402,doi:10.1063/1.4976974}.

In future studies we will extend the proposed finite size corrections to the study of excited states and metallic systems.
We note that excited states and spectral functions can be calculated in the framework of equation of motion coupled cluster theory
for solids, yielding excited state structure factors that are expected to exhibit similar finite size errors~\cite{PhysRevB.93.235139}.
In metallic systems the structure factor is still algebraic around ${\bf G}=0$.
We are therefore confident that the proposed methods can also be transfered to the study of such systems
and we expect the outlined twist averaging methodology
will be of significant importance when approaching the thermodynamic limit.
We note, however, that the perturbative triples (T) contribution requires methodological
improvements when applied to metals. 

%
%

\emph{Acknowledgements.} -- 
This project has received funding from the European Research Council (ERC) under the European Union’s Horizon 2020 research and innovation program (grant agreement No 715594).
The computational results presented have been achieved in part using the Vienna Scientific Cluster (VSC).

\bibliography{cite_grueneis,cite_tsatsoulis}

\end{document}